\documentclass[a4paper,prl,showpacs,twocolumn,amssymb]{revtex4}

\usepackage{bm} 
\usepackage{times}
\usepackage{graphicx}

\newcommand{\Sc}{Schr\"o\-din\-ger }
\newcommand{\diag}{\mbox{\rm diag}}

\begin{document}

\title{Reconstructing the nucleon-nucleon potential
by a new coupled-channel inversion method}
\author{Andrey Pupasov}
\author{Boris F. Samsonov}
\affiliation{Physics Department, Tomsk State University,
36 Lenin Avenue, 634050 Tomsk, Russia}
\author{Jean-Marc Sparenberg}
\author{Daniel Baye}
\affiliation{Universit\'e Libre de Bruxelles (ULB),
Physique Nucl\'eaire et Physique Quantique,
CP 229, B 1050 Bruxelles, Belgium}
\date{\today}

\begin{abstract}
A second-order supersymmetric transformation is presented,
for the two-channel Schr\"odinger equation with equal thresholds.
It adds a Breit-Wigner term to the mixing parameter,
 without modifying the eigenphase shifts,
and modifies the potential matrix analytically.
The iteration of a few such transformations allows a precise
fit of realistic mixing parameters
in terms of a Pad\'e expansion of both the scattering matrix
 and the effective-range function.
The method is applied to build an exactly-solvable potential
 for the neutron-proton $^3S_1$-$^3D_1$
case.
\end{abstract}


\maketitle

In quantum scattering theory,
the fundamental inverse problem consists in deducing the interaction potential
between two colliding particles
from their experimental elastic-scattering cross sections \cite{chadan:89}.
These cross sections have first to be parametrized
in terms of energy-dependent partial-wave phase shifts or scattering matrices.
For a central interaction $V(r)$,
the partial waves decouple and a sequence
 of single-channel inverse problems have to be solved.
For more complicated interactions,
like the tensor interaction in nuclear physics,
partial waves may be coupled and matrix potentials have
to be constructed from coupled-channel
scattering matrices.

A formal solution to these inverse problems can be written
in terms of integral equations
\cite{chadan:89}.
In practical applications,
experimental data turn out to be precisely parametrizable
in terms of separable kernels for these equations.
These correspond to scattering matrices that are rational
functions of the wave number.
The integral equations can then be solved analytically
and the corresponding potentials are also expressed in a separable form.
This procedure applies to both the single- and coupled-channel cases
\cite{newton:57,kohlhoff:93}.
In the single-channel case,
the same potentials can be more efficiently constructed
with the help of supersymmetric quantum mechanics (SUSYQM)
\cite{nieto:84,sukumar:85c},
which directly relates the potentials to their scattering-matrix poles.
Moreover, a small number of poles is in general sufficient
to fit experimental data on the whole elastic-scattering energy range;
this is in particular the case for the neutron-proton singlet (spin 0) channels
\cite{sparenberg:97a,samsonov:03},
that decouple because of the vanishing tensor interaction.

The present Letter aims at extending this very efficient sin\-gle-channel method to
the two-channel case without threshold difference,
and at applying it to the neutron-proton triplet (spin 1) coupled channels.
This system was studied in the framework
of the integral-equation method by Newton and Fulton
\cite{newton:57}, at low energy, and by Kohlhoff and von Geramb \cite{kohlhoff:93},
on the whole elastic-scattering range.
The present paper subsumes these works,
by separating the effect of the coupling between channels in the inversion procedure,
by parametrizing the data with a minimal number of poles,
and by deriving simple expressions for the corresponding matrix potential.

We consider two channels with equal thresholds and angular momenta $l$ and $l+2$.
The scattering matrix is parametrized by the eigenphase shifts $\delta_1$, $\delta_2$
and the mixing parameter $\epsilon$,
\begin{equation}
S(k)=R[\epsilon(k)]{\rm\ diag}\left(e^{2i\delta_1(k)},
e^{2i\delta_2(k)}\right)R^T[\epsilon(k)]\,.
\end{equation}
Here, $k$ is the (complex-valued in general) wave number,
corresponding to the center-of-mass energy $E=k^2$ in reduced units,
and $R$ is an orthogonal matrix
\begin{equation}\label{rot-def}
\qquad R(\epsilon)= \left(
\begin{array}{cc}
 \cos\epsilon & \sin\epsilon
\\ -\sin\epsilon &\cos\epsilon
\end{array}
\right).
\end{equation}
We aim at building a symmetric interaction matrix $V$ with entries
$V_{11}$, $V_{12}$, $V_{22}$,
by inversion of the scattering data $\delta_1,\delta_2,\epsilon$.
In Ref.~\cite{pupasov:10}, we show how
to split this two-channel inverse problem into two independent parts:
(i)~fitting the eigenphase shifts independently for each channel,
with the standard single-channel method
\cite{sukumar:85c,sparenberg:97a,samsonov:03}, and
(ii)~fitting the mixing parameter,
with a new type of eigenphase preserving (EPP) transformation.
In Ref.\ \cite{pupasov:10},
such a transformation is introduced but it is restricted
to a mixing parameter which is an {\em odd} function of the energy,
as in Ref.\ \cite{newton:57}.
Here, we overcome this restriction,
which allows us to fit more realistic mixing parameters.

Let ${H_{0}=-I_2\, {d^2/dr^2} + V_{0}}$ be an initial $2\times 2$ matrix Hamiltonian,
where $I_2$ is the identity matrix and the potential $V_0(r)$ is symmetric.
With two successive SUSYQM transformations,
this Hamiltonian can be transformed into another Hamiltonian $H_2$,
which has the same eigenphase shifts as $H_0$ but a different mixing parameter.
Equivalently,
this can be achieved with a two-fold EPP transformation
defined by the intertwining relation $LH_0=H_2L$,
where $L$ is a second-order differential matrix operator,
called  transformation operator,
which maps solutions of both Hamiltonians as
$\Psi_2(k,r)=L\Psi_0(k,r)$.
Operator $L$ and its adjoint $L^\dagger$ obey the factorization properties
$L^\dagger L=(H_0-E_1)(H_0-E_1^*)$,  $L L^\dagger=(H_2-E_1)(H_2-E_1^*)$,
where $E_1 = k_1^2 = E_{R1} + i E_{I1}$ and $E_1^* = E_{R1} - i E_{I1}$
are mutually conjugate factorization constants.
General properties and different equivalent expressions
for $L$ can be found in Ref.~\cite{pupasov:10}.

The transformed potential reads
\begin{equation}
V_2(r)=V_0(r)-2 W_2'(r).
\label{mat-pot}
\end{equation}
The matrix $W_2$ is expressed in terms of a complex matrix $U$,
made up of two vector solutions of the coupled-channel \Sc equation,
$-U''+ V_{0} U = E_1 U$, as \cite{pupasov:10}
\begin{equation}
W_2(r) = -E_{I1} \left\{{\rm Im}[ U'(r) U^{-1}(r)] \right\}^{-1}.
\end{equation}
Potential $V_2$ is real, symmetric, and re\-gu\-lar
when the first vector of $U$ is regular for $r\rightarrow 0$
and its asymptotic behavior reads
\begin{equation}
U(r\rightarrow\infty) \rightarrow
\left(\begin{array}{cc}
       e^{-ik_1 r} & -i e^{i k_1 r} \\
      i e^{-i k_1 r} & e^{i k_1 r}
      \end{array}
\right).
\end{equation}

As announced, for this EPP transformation,
the eigenphase shifts of the transformed scattering matrix $S_2$ coincide
with the initial ones: $\delta_{c;2}(k)=\delta_{c;0}(k)$ for channel $c=1,2$.
On the other hand,
the mixing parameter is modified as
\begin{equation}\label{ppt-mpsd-gen}
\epsilon_2(k)=\epsilon_0(k)+\arctan \frac{E_{I1}}{E_{R1}-k^2},
\end{equation}
which corresponds to additional $S$-matrix poles in $k^2=E_{R1}\pm i E_{I1}$.
In Ref.~\cite{pupasov:10},
this result was proved in the case of purely imaginary poles,
$E_{R1}=0$.
We made this choice because a mixing parameter should vanish at zero energy
(up to an unimportant integer multiple of $\pi/2$)
for the potential to be short ranged (except
for the $r^{-2}$ centrifugal diagonal term).
For $E_{R1}\ne 0$, all calculations of Ref.\ \cite{pupasov:10} are actually valid
but lead to a non-vanishing mixing parameter,
$\epsilon_2(0) = \arctan(E_{I1}/E_{R1}) \equiv \alpha_1$,
and to $r^{-2}$ off-diagonal terms in the potential.
Here, we solve these problems by applying an energy-independent rotation
 after the EPP transformation:
the $S$ matrix $S_2$ transforms into $\bar{S}_2=R^T(\alpha_1) S_2 R(\alpha_1)$,
which shifts the mixing parameter by a constant value,
$\bar\epsilon_2(k) = \epsilon_2(k)-\alpha_1$,
whence $\bar\epsilon_2(0)=0$.
Meanwhile, $H_2$ transforms into $\bar H_2=R^T(\alpha_1)H_2R(\alpha_1)$,
which has no $r^{-2}$ off-diagonal term.

Iterating $P$ times such EPP transformations, with
energies $E_j=E_{Rj} \pm i E_{Ij}$, $j=1,\ldots,P$,
one builds a chain of Hamiltonians, $H_0\to H_2\to$ $\ldots\to
H_{2P}$.
The final potential $V_{2P}$ corresponds to the eigenphase equivalent $S$ matrix
\begin{equation}
S_{2P}(k) = R(\bar{\epsilon}_{2P}) \diag\left( e^{2i\delta_{1;0}},
 e^{2i\delta_{2;0}}\right) R^T({\bar{\epsilon}_{2P}}),
\end{equation}
with $\bar{\epsilon}_{2P}(k)=\epsilon_{2P}(k)-\epsilon_{2P}(0)$ and
\begin{equation} \label{eps2P}
\epsilon_{2P}(k) =
\epsilon_0(k)+\sum\limits_{j=1}^P\arctan \frac{E_{Ij}}{E_{Rj}-k^2}.
\end{equation}
As above, to get a vanishing zero-energy mixing parameter,
we perform a compensation rotation of angle
$-\epsilon_{2P}(0)=-\sum_{j=1}^P \alpha_j$,
with $\alpha_j \equiv \arctan(E_{Ij}/E_{Rj})$, $j=1,\dots,P$,
on both the scattering matrix and the Hamiltonian $H_{2P}$.

The sum in parametrization (\ref{eps2P}) is flexible enough
to fit realistic mixing parameters,
even with small values of $P$.
This can be explained with an effective-range expansion:
the elements of matrix (\ref{rot-def}) are analytical at the origin
and can be Taylor expanded as a function of the energy \cite{delves:58}.
Consequently, $\tan \epsilon$ is well approximated by a Pad\'e expansion,
which is exactly what our EPP transformations lead to:
for $\epsilon_0=0$, $\tan \bar{\epsilon}_{2P}$, with $2P$ arbitrary parameters,
is the most general Pad\'e approximant of order $[P/P]$ in energy,
vanishing at the origin.

This is very analog to the standard single-channel SUSYQM inversion
\cite{sukumar:85c,sparenberg:97a,samsonov:03},
where Pad\'e approximants also occur.
There, the one-channel $S$ matrix is expanded in terms of its poles $i\kappa_m$
in the complex wave-number plane as \cite{sparenberg:97a,samsonov:03}
\begin{equation}
S_{l}(k) = e^{2i\delta_l(k)}
= \prod_{m=1}^{M_{l}} \frac{i \kappa_m+k}{i \kappa_m-k}
\quad(M_l\ge 2l+1),
\label{Sprod}
\end{equation}
which corresponds to the single-channel phase shift
$\delta_{l}(k)=-\sum_{m=1}^{M_{l}} \arctan (k / \kappa_m)$.
The poles are generally restricted to the imaginary axis,
except for resonances,
and have to be chosen so that the effective-range function
\begin{eqnarray}
K_l(k^2) = k^{2l+1} \cot \delta_l(k) = ik^{2l+1} \frac{S_l(k)+1}{S_l(k)-1}
\label{KS}
\end{eqnarray}
does not vanish at zero energy,
which corresponds to a finite scattering length.
At higher orders, $K_l$ is usually expanded as a Taylor series,
but this expansion often breaks down at high energy.
The $S$ matrix (\ref{Sprod}), in contrast,
leads to a Pad\'e approximant for $K_l$,
of order $\left[\frac{M_l}{2}/\frac{M_l}{2}-l-1\right]$ in energy for even $M_l$
and of order $\left[\frac{M_l-1}{2}/\frac{M_l-1}{2}-l\right]$ for odd $M_l$.
Such approximants are valid at any energy and are able to
 fit experimental phase shifts with high precision,
even for a limited order.

Expansion (\ref{Sprod}) has the additional advantage that
the corresponding single-channel potential is known analytically
\cite{sparenberg:97a},
\begin{equation}\label{susy-pot}
 V_{l;M_l}(r)  =  \frac{l(l+1)}{r^2}-2\frac{d^2}{dr^2} \ln W[u_1,\ldots,u_{M_l}]\,,
\end{equation}
where $W$ is a Wronskian determinant of functions $u_m(r)$.
These functions are solutions of the free radial \Sc equation,
$-u_m'' + l(l+1)r^{-2} u_m = E_m u_m$,
at energies $E_m=-\kappa_m^2$ given by the $S$-matrix poles.

Combining this single-channel inversion technique with the
 EPP transformations provides a complete
coupled-channel inversion scheme.
First, experimental eigenphase shifts are inverted separately for each channel,
i.e., they are fitted with Eq.\ (\ref{Sprod})
and the corresponding potential is constructed
with Eq.\ (\ref{susy-pot}).
This leads to diagonal potential and scattering matrices,
which can then be used as a starting point for EPP transformations.
The factorization energies of these transformations are chosen
by fitting the experimental mixing parameter with Eq.\ (\ref{eps2P})
and the corresponding coupled-channel potential
is finally constructed by iterated application of Eq.\ (\ref{mat-pot}),
followed by a compensation constant rotation.
The final potential is thus directly built from its scattering-matrix poles,
which are either associated to the eigenchannels or to the coupling.

\begin{figure}
\scalebox{0.77}{\includegraphics{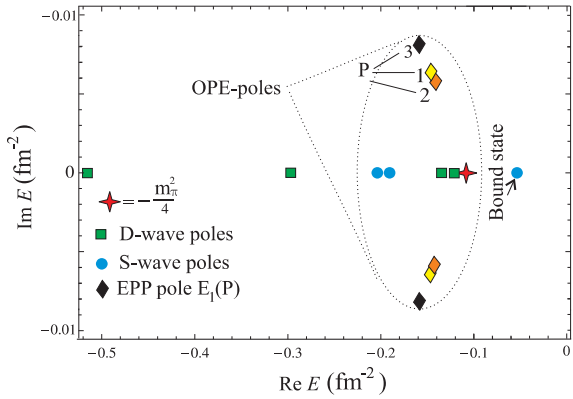}}
\caption{(Color online) Poles of
the neutron-proton $^3S_1$-$^3D_1$ scattering matrix close to the
origin (note the different scale for the imaginary energy axis).
\label{fig-poles-e}}
\end{figure}

Let us now consider one of the most important coupled-channel inverse problems,
i.e., the inversion of the neutron-proton $^3S_1$-$^3D_1$ scattering matrix.
To test our method,
we use the scattering matrix of the Reid93 potential \cite{stoks:94} as input data.
Using Eq.\ (\ref{Sprod}) with 5 terms for both the $s$ and $d$ waves ($M_s=M_d=5$),
we find the following ten factorization constants
$s_m=0.23154$, $-0.45146$, 0.43654, 1.6818, 2.3106 fm$^{-1}$,
$d_m=-0.36719$, $-0.54420$, 0.34828, 0.71766, 3.3758 fm$^{-1}$,
which fit the neutron-proton
 $^3S_1$ and $^3D_1$ phase shifts, respectively.
Note that $s_1=0.23154$~fm$^{-1}$ is fixed to reproduce
the deuteron binding energy $B_d=2.2245$ MeV.
The $S$-matrix poles $-s_m^2$ and $-d_m^2$
closest to the origin are represented
in the complex energy plane in Fig.\ \ref{fig-poles-e}.

\begin{figure*}
\scalebox{0.67}{\includegraphics{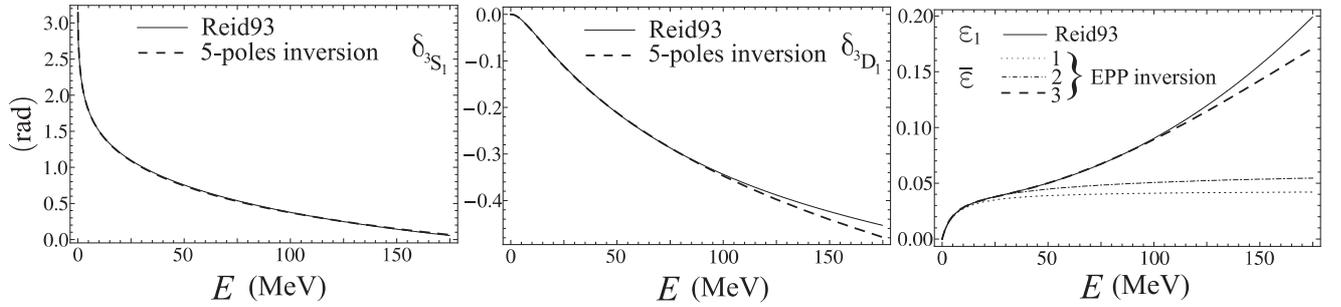}}
\caption{Eigenphase shifts $\delta_{^3S_1}$, $\delta_{^3D_1}$
and mixing parameter $\epsilon_1$
of the $^3S_1$-$^3D_1$ channels for neutron-proton scattering.
\label{fig-4mp}}
\end{figure*}

\begin{figure*}
\scalebox{0.67}{\includegraphics{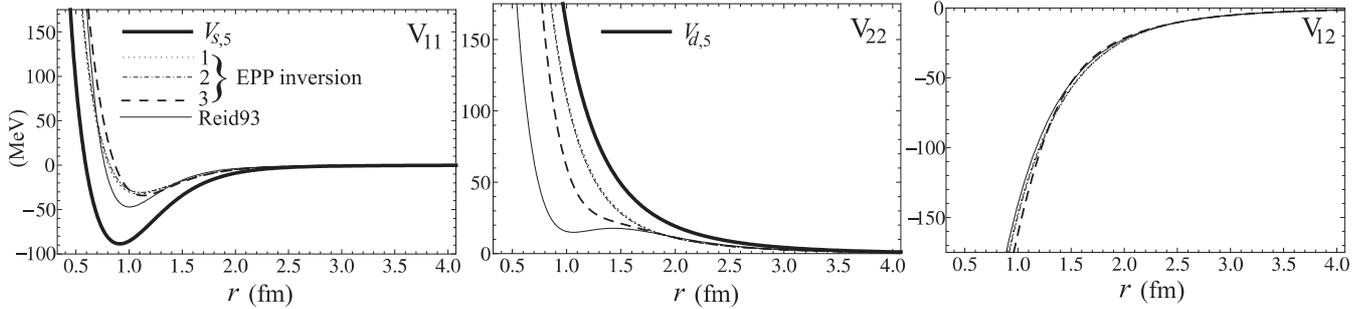}}
\caption{Reid93 potential and its inversion potentials
for the neutron-proton $^3S_1$-$^3D_1$ channels.
\label{fig-reidpt-e}}
\end{figure*}

\begin{figure}
\scalebox{0.36}{\includegraphics{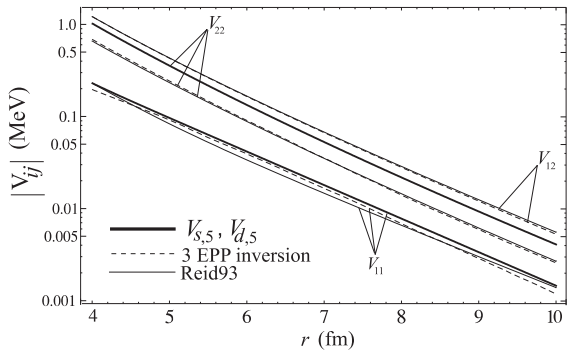}}
\caption{Asymptotic behavior of the Reid93 and inversion potentials.
\label{fig-2log}}
\end{figure}

In Fig.\ \ref{fig-4mp}, we compare
the eigenphase shifts of the Reid93 potential and our fit.
The corresponding potentials
$V_{s;5}$ and $ V_{d;5}$ are determined by Eq.\ (\ref{susy-pot}) with the following
sets of solutions:
$u_{s;m}=\exp(s_{m} r)$, $u_{s;n}=\sinh(s_{n} r)$,
and $u_{d;m}=\left[3 d_{m} r
\cosh( d_{m}r) -(3+d_{m}^2 r^2) \sinh(d_{m}r)\right]/r^2$,
$u_{d;n}={\rm e}^{-d_{n}r}\left[1+3/(d_{n}r)+3/(d_{n}r)^2\right]$,
$m=1,2$, $n=3,4,5$.
After extracting the centrifugal term from the second channel,
we plot these potentials in Fig.\ \ref{fig-reidpt-e},
as well as their asymptotic behavior in logarithmic scale in Fig.\ \ref{fig-2log}.
At large distances, the one-pion-exchange (OPE) short-range behavior
of the Reid93 potential,
with a pion mass of $m_\pi\approx 0.684$ fm$^{-1}$,
is clearly seen.
Both inversion potentials are also short ranged.
For the $s$-wave potential, this is surprising at first sight:
in general, for a SUSYQM inversion,
the leading contribution to the potential asymptotics comes from the pole
$i\kappa_m$
with $\kappa_m>0$ closest to the origin,
and behaves like $\exp(-2 \kappa_m r)$.
In the present case,
this is the bound-state pole $i s_1$,
which lies closer to the origin than the OPE cut,
and one would expect the inversion potential to decrease slower than the OPE
potential at large distances.
We solve this problem here by exploiting the degree of freedom provided
by the presence of a bound state:
in our inversion technique,
the bound-state asymptotic normalization constant (ANC) can be chosen arbitrarily
without affecting neither the phase shift nor the binding energy \cite{baye:87b}.
For a particular value of the ANC,
related to the residue of the $S$-matrix bound-state pole \cite{sparenberg:10},
the $\exp(-2 s_1 r)$ potential tail vanishes;
for our single-channel $S$-matrix parametrization,
and thus for potential $V_{s;5}$,
this ANC has the value $A_{s;5}=0.8854$ fm$^{-1/2}$.

Despite this satisfactory short-range behavior,
potentials $V_{s;5}$ and $V_{d;5}$
differ from  the Reid93 diagonal potentials:
Figs.\ \ref{fig-reidpt-e} and \ref{fig-2log} show that
$V_{s;5}$ is more attractive,
whereas $V_{d;5}$ is more repulsive.
This confirms that a coupling is necessary to recover the original potential.
We thus transform the diagonal potential $V_0={\rm diag}(V_{s;5},V_{d;5})$
into a coupled one by using EPP transformations.
The $P$ complex energies $E_j$ are chosen
to reproduce physical properties of the Reid93 potential.
For positive energies, we fit the mixing parameter,
while for the negative bound-state energy,
we fit the ratio of ANCs for the $d$- and $s$-wave components,
$\eta=A_d/A_s$,
to its value for the Reid93 potential, $\eta=0.0251$ \cite{stoks:94}.
These ANCs are related to the residue of the two-channel scattering
matrix at the bound-state pole \cite{stoks:88} and read
$A_s=A_{s;5} \cos \epsilon(-B_d)$, $A_d=-A_{s;5} \sin \epsilon(-B_d)$.
Hence, one has to satisfy $\tan \epsilon(-B_d)=-0.0251$.
For the Reid93 potential,
the mixing parameter is not an odd function of the energy,
which confirms the need for the general EPP transformations introduced above.
A single transformation ($P=1$) provides two free parameters,
$E_{R1}$ and $E_{I1}$;
we fix them to reproduce $\eta$ and the slope of the mixing parameter at zero energy,
$\epsilon'(0)=0.297$ MeV$^{-1}$.
The values of $E_{R1}$, $E_{I1}$ compatible with these constraints
are given in Table \ref{tab-epp} for $P=1$.
The mixing parameter $\bar\epsilon_2$ is shown in Fig.\ \ref{fig-4mp} by dots.
It is seen that this model provides a good low-energy fit,
in the spirit of the Newton-Fulton potential \cite{newton:57};
however, the present potential is much simpler and more general,
as it allows a mixing parameter which is not an odd function of the energy.

\begin{table}\caption{\label{tab-epp}
Real $E_{Rj}$ and imaginary $E_{Ij}$ parts of the $P$ energies $E_j$
entering the EPP transformations ($j=1,\dots,P$).}
\begin{ruledtabular}
\begin{tabular}{lllllll}
$P,j$                    & 1,1 & 2,1 & 2,2 & 3,1 & 3,2 & 3,3 \\
\hline
$-E_{Rj}$ (fm$^{-2}$) & $0.1463$    &  $0.1430$
& $2.191$    & $0.1590$   & $4.004$ & 0 \\
$E_{Ij}$ ($10^{-3}$ fm$^{-2}$) & $6.372$  & $5.864$
 & $49.85$  & $8.153$ & $-3107$ & $9443$ \\
\end{tabular}
\end{ruledtabular}
\end{table}

\begin{table}\caption{ \small \label{tab-dp}
Deuteron properties ($s$-wave ANC, $d$-wave probability)
for the Reid93 and the $P$-transformation inversion potentials.}
\begin{ruledtabular}
\begin{tabular}{llllll}
Property    & Reid93           &           $P=1$ &  $P=2$            &  $P=3$\\
\hline
$A_s$ (fm$^{-1/2}$) & 0.8853 & 0.8851 & 0.8851 & 0.8851 \\
$P_d$ (\%)     & 5.699            &    5.27        & 5.39             &  5.73  \\
\end{tabular}
\end{ruledtabular}
\end{table}

Fitting the mixing parameter at higher energies requires additional free parameters.
Therefore, we iterate the EPP transformations.
In Table \ref{tab-epp}, we list the positions of the $S$-matrix poles
for $P=1,2,3$ EPP transformations.
The corresponding mixing parameters are plotted in Fig.\ \ref{fig-4mp},
which shows a good improvement with increasing $P$.
The effective potentials are shown in Fig.\ \ref{fig-reidpt-e}
(the $P=1$ and $P=2$ potentials
are hardly distinguishable at the scale of the figure).
They still display some differences with the Reid93 potential,
due to their different high-energy $S$-matrix,
but Fig.\ \ref{fig-2log} shows that,
asymptotically, our $P=3$ potential is very close to the Reid93 potential.
In Fig.\ \ref{fig-poles-e},
we show that some of the $S$-matrix poles
are concentrated near $E=-m_\pi^2/4$ and are nearly independent of $P$.
These poles can be associated  with the OPE contribution to the $np$ interaction.
Other deuteron properties
are given in Table \ref{tab-dp} for the inversion potentials;
a quick convergence towards the Reid93 values is also seen.

In conclusion, we have presented in this Letter an optimal
cou\-pled-channel inversion algorithm,
combining standard single-channel SUSYQM
 transformations with coupled-channel EPP transformations
that display the remarkable feature of modifying the mixing parameter
without affecting the eigenphase shifts.
The method is developed for two channels with equal thresholds
and we plan to study its applicability to more general cases.
In the neutron-proton $^3S_1$-$^3D_1$ case,
the method leads to a simple potential,
directly related to the position of its scattering-matrix poles.
This very efficient parametrization could also be used
in a phase-shift analysis of experimental data,
directly providing the corresponding potential.

\acknowledgments{
This text presents research results
of the Belgian Research Initiative on eXotic nuclei (BriX),
program P6/23 on interuniversity attraction
poles of the Belgian Federal Science Policy Office.
The work is supported by the  Russian government
 under contracts Nos.\ 02.740.11.0238, P1337 and P2596.
}


\end{document}